\documentclass[aps,prd,10pt,
               preprintnumbers,numbers,sort&compress,nofootinbib,
               showpacs]{revtex4}

\usepackage{amsmath}
\usepackage{amssymb}
\usepackage{hyperref}

\newcommand{\exclude}[1]{}

\newcommand{\beq}{\begin{equation}}
\newcommand{\eeq}{\end{equation}}
\newcommand{\bea}{\begin{eqnarray}}
\newcommand{\eea}{\end{eqnarray}}
\newcommand{\nn}{\nonumber}

\def\la{\langle }
\def\ra{ \rangle }

\def\la{\langle 0 |}
\def\ra{ | 0 \rangle }

\newcommand{\D}{{\cal D}}

\newcommand{\junk}[1]{}

\begin{document}

 \title{  Phase Transitions, $\theta$ Behavior and Instantons in QCD and its Holographic Model 
} 
  \author{ Andrei Parnachev}
   \affiliation {C.N. Yang Institute for Theoretical Physics, University of Stony Brook, Stony Brook NY 11794-3840, USA}

\author{Ariel R. Zhitnitsky}
\affiliation{Department of Physics and Astronomy, University of
  British Columbia, Vancouver,  Canada}

\begin{abstract}
In the holographic model of QCD, $\theta$ dependence sharply changes
at the point of confinement-deconfinement phase transition.
In large $N$ QCD such a change in $\theta$ behavior can be related
to the breakdown of the instanton expansion at some critical temperature $T_c$.
Associating this temperature with confinement-deconfinement phase 
transition leads to the description of the latter in terms of 
dissociation of instantons into the fractionally charged instanton quarks.
To elucidate this picture, we introduce the nonvanishing chiral condensate 
in the deconfining phase and assume a specific lagrangian for the $\eta'$ field
in the confining phase.
In the resulting picture the high temperature phase of the theory consists of the dilute gas of instantons, while
the low temperature phase is described in terms of freely moving  fractional instanton quarks.
\end{abstract}
 
\preprint{NSF-KITP-08-76}

\maketitle

\section{ Introduction}  
Color confinement, spontaneous breaking of chiral symmetry, 
the $U(1)$ problem and the $\theta$ dependence are some of the most
interesting questions in $QCD$.   At the end of the 1970's A. M. Polyakov \cite{Po77}
demonstrated charge confinement in $QED_3$.  This was the first example
where nontrivial dynamics was shown to be a key ingredient for
confinement: 
Instantons  (monopoles in 3d) play a crucial role
in  the dynamics of confinement in $QED_3$.   
 Instantons in four dimensional QCD were discovered  more than 30  years ago
\cite{Belavin:1975fg}. However, their role in   $QCD_4$ is  still  not  well understood 
  due to the divergence  of the instanton density for 
large size instantons in the confining phase. Soon after, 't Hooft and Mandelstam
\cite{Hooft} suggested a qualitative picture of how confinement could
occur in $QCD_4$.  The key point in  the 't Hooft - Mandelstam approach
is the assumption that dynamical monopoles exist and Bose
condense. Many papers have been written on this subject since the
original formulation \cite{Hooft}; however, the main questions, such
as, ``What are these monopoles?''; ``How do they appear in the gauge
theories without Higgs fields?''; ``How do they interact?''; `` What is the relation (if any)
between the   't Hooft - Mandelstam  monopoles and instantons?" , are
still not understood (for a   review see \cite{Hooft1}).

We  re-consider these issues  from a slightly different angle
 by analyzing  phase transitions as a function of  temperature, $T$, at nonzero  $\theta$ parameter.
  We study  the evolution of the most important field configurations
as the phase transition line is crossed.
Indeed, understanding  $\theta$ dependence gives a very good idea 
about the dynamics of the most important color  fluctuations with nontrivial topology.   
On the other hand, we will see that
$\theta$ dependence can be studied using an effective lagrangian approach with  color singlet degrees of freedom. 

The main result of our \exclude{study} work can be formulated as follows. 
In the holographic model of QCD \cite{Witten:1998zw,Sakai:2004cn}
confinement -deconfinement phase transition 
happens precisely at the value of temperature $T=T_c$ where $\theta$ dependence 
experiences a sudden change in behavior from $\theta/N$ in the low-temperature (confining)
phase to $e^{-N} \cos\theta$ in the high temperature (deconfining) phase \cite{Bergman:2006xn}.
Consider now QCD with large number of colors $N$.
For very high temperatures $T>>T_c\sim \Lambda_{QCD}$ the typical size of instantons
is very small and the instanton gas is dilute 
with density of order $e^{-N}$. Calculations in this region are
under complete field theoretic control and the vacuum energy 
behaves like $e^{-N} \cos\theta$. 
As the temperature is lowered to be of order $T_c$ the average instanton grows in size
and the perturbative expansion around the instanton field 
configuration becomes unreliable.
However, we argue, based on some reasonable assumptions, that in the large $N$ limit
the average distance between the instantons remains much larger then
the instanton size all the way down to the critical value of temperature $T=T_c$.
Below $T_c$  the instanton expansion 
breaks down and a consistent field theoretic  calculation  with non-overlapping instantons is no longer  possible. 
It is then natural to assume that at $T=T_c$ there is a sharp transition in
$\theta$ behavior, which 
can be associated with confinement-deconfinement transition,
just as in the holographic model.
The value of $T_c$ can be estimated by a one-instanton calculation.

To elucidate the physics of the transition we consider 
a model where  the chiral  condensate \exclude{develops} does not vanish in the deconfining phase. 
The holographic model of QCD is a good example where this phenomenon occurs.
On the field theoretic side this can be achieved by coupling fundamental matter
to the hidden gauge group whose dynamically generated energy scale is 
higher than that of QCD.
In the presence of nonvanishing chiral condensate, 
its phase $\varphi$ (which can be canonically normalized to yield $\eta'$)
is a perfect  probe  of    the topological 
charges of  the constituents  on both sides of the phase transition line.   
This is a consequence of  uniqueness of  $\eta'$ meson: it  always enters the system  
in combination  $(\theta-\varphi)$ irrespectively whether it is in the confining or 
the deconfining phase. One should remark here that our results
do not really depend on 
the value of $N_f$ as long as $N_f\ll N$.

The plan of the paper is as follows.
We start in  Section II  by reviewing  recent work on the holographic model of QCD
where we 
note that $\theta$ behavior sharply changes 
at the point of confinement-deconfinement 
phase transition.
 We return from holographic model to quantum field theory \exclude{description}  in section III
 where we argue that instanton expansion breaks down sharply at some critical temperature
$T_c$ and estimate its value
in terms of $\Lambda_{QCD}$.
 Sections IV and V 
 are devoted to the physical interpretation of the phase transition. 
 Here we attempt to answer  the following question:
 what happens to the well-defined objects (instantons) as the phase transition
 line at $T=T_c$ is crossed from above. 
We assume a certain lagrangian for the low temperature phase and show,
that under this assumption,  
instantons do not completely  disappear from the system 
but rather
dissociate  into the instanton quarks\footnote{ Instanton quarks originally appeared in 2d models.
  Namely, using an exact accounting and resummation
of the $n$-instanton solutions in $2d~CP^{N-1}$ models,
 the original problem of a statistical instanton
 ensemble  was   mapped unto a $2d$-Coulomb Gas (CG) 
system of pseudo-particles 
with fractional topological charges $\sim 1/N$ \cite{Fateev}.    
This picture leads to the 
elegant explanation of the confinement phase and other important properties of
the $2d~CP^{N-1}$ models \cite{Fateev}.  
Unfortunately,  similar calculations  in  $4d$ gauge theories
is proven to be much more difficult  to carry out  \cite{Belavin}. },
 the objects with fractional topological charges $\pm 1/N$
which become the dominant quasi-particles.
At non-zero temperatures $T<T_c$, 
the instanton quarks carry, along with fractional topological charges,  
    the fractional magnetic charges  which  makes them the perfect candidates
   to serve as  the dynamical  magnetic  monopoles, the crucial element  
   of  the standard  't Hooft and Mandelstam picture for the confinement \cite{Hooft}, 
see the end of section V for details.
In section VI we formulate  our main results and discuss future directions.

\section{Holographic model of QCD}
In this section we consider physics of holographic model of large $N$ QCD with $N_f\ll N$ 
flavors \cite{Witten:1998zw,Sakai:2004cn}. 
This section is the review of previous work; our main point here
is the observation that $\theta$ dependence changes once we go from the 
confining to the deconfining phase.
The holographic model of QCD is realized by placing $N_f$ $D8-\bar D8$
pairs in the background created by $N$ $D4$ branes.
In the weak coupling regime the D4 branes span $x^0\ldots x^4$ coordinates,
while  $D8-\bar D8$ branes are pointlike in the $x^4$ direction and span the
rest of spacetime.
One of the directions (denoted by $x_4$ below) along the $D4$ branes is compactified on a circle
of radius $R_4$ (which sets the scale of the glueball masses) with antiperiodic boundary conditions for fermions.
The value of the asymptotic separation between the $D8$ and $\bar D8$ branes, denoted by  $L$, is a parameter of 
the brane construction 
along with $R_4,N,N_f$ and string coupling and length $g_s$ and $l_s$
(which will be set to unity in the rest of the paper).
It will be convenient to introduce the five-dimensional t'Hooft coupling $\lambda_5=g_s N l_s$
and its four-dimensional counterpart defined at Kaluza-Klein scale $\lambda_4=\lambda_5/R_4$.
In the limit $\lambda_4\ll 1$, $\Lambda_{QCD}\ll 1/R_4$ and hence the theory approximates
QCD pretty well.
String theory is solvable in the opposite regime, $\lambda_4\gg 1$, where there is no clear
separation between the QCD scale and the supergravity/DBI dynamics.
It is this regime that we consider below, in the hope of drawing some qualitative
lessons. 

For $\lambda_4\gg 1$ the rules of gauge/string duality \cite{Maldacena:1997re,Witten:1998qj,Gubser:1998bc} instruct us to 
pass to the metric which is the product of the D4 branes backreaction:
\beq
  ds^2=\left(\frac{U}{R}\right)^\frac{3}{2} \left( (dx_\mu)^2 + f(U) (dx^4)^2\right)+
       \left(\frac{U}{R}\right)^{-\frac{3}{2}} \left( \frac{dU^2}{f(U)}+U^2 d\Omega_4^2\right)
\label{metric}
\eeq
where $f(U)=1-U_K^3/U^3$ and the $U$ coordinate is bounded from below by $U_K$.
The $(U,x^4)$ are the analogs of polar coordinates on the plane, which is ensured
by the relation
\beq
  2\pi R_4= \frac{4\pi}{3}\left(\frac{R^3}{U_K}\right)^\frac{1}{2}=
              \frac{4\pi}{3}\left(\frac{\pi\lambda_5}{U_K}\right)^\frac{1}{2}
\label{reluk}
\eeq
where in the second equality we used the relation between the
t'Hooft coupling and the curvature scale of the space (\ref{metric}).

As explained in \cite{Witten:1998uka}, the inclusion of the $\theta$ angle
in this model corresponds to having nonvanishing integral of RR one-form over the $x^4$ circle
\beq
     \int_{S^1} C_1= \int_D F_2=\theta\;mod \;2\pi k;\qquad F_2=d C_1
\label{thetahqcd}
\eeq
where $k$ is an integer number and the integral is over the $S^1$ parameterized
by $x^4\in[0,2\pi R_4]$.
In the first equality we used the fact that the $(U,x^4)$ space has the disk topology
and Stokes theorem. 
One can solve the equation of motion for $F_2$ without taking back-reaction into
account (which is justified as long as  $N_f\ll N$) and substitute into
the action; the result for the vacuum energy at small $\theta$ is \cite{Witten:1998uka}
\beq
  E_{vac}\approx {\chi_g\over2} \theta^2
\label{evachqcd}
\eeq
where $\chi_g\sim {\cal O}(1)$ is the topological succeptibility.
The addition of fundamental matter results \cite{Sakai:2004cn} in the effective lagrangian 
consistent with Veneziano-Witten formula for the $\eta'$ mass:\footnote{ Earlier works on
the holographic derivation of $\eta'$ lagrangian include \cite{Armoni:2004dc} and \cite{Barbon:2004dq}.
Theta dependence in the holographic models has been also considered in \cite{Ka,Kb}.}
\beq
  {\cal L}_{eff}= \frac{1}{2}(\partial_\mu\eta')^2+ 
            \frac{N^2}{2}\chi_g \left(\frac{\theta}{N}+\frac{1}{N}\frac{\eta'}{f_{\eta'}}\right)^2 ,
\label{effetap}
\eeq
where we included some numerical factors such as $\sqrt{2}$ and $\sqrt{N_f}$ into the definition of 
$f_{\eta'}$ to simplify notations in the following sections.
This result is not significantly changed when the finite temperature is introduced, 
as long as the theory is in the confining phase and the topology of the space remains the same.
Eq. (\ref{evachqcd}) is consistent with the fact that in the confining phase
physics is expected to depend on $\theta$ via the combination $\theta/N$\cite{Witten:1998uka},
\beq
   E_{vac}=N^2 \min_k \; h\left(\frac{\theta+2\pi k}{N}\right),   
\label{evacmv}
\eeq
where $h(x)$ is some function which satisfies $h(0)=h'(0)=0$.
 Eq.(\ref{evacmv})  can also  be understood   from QFT viewpoint  for finite $N$ as a result of   summation  over different branches in pure $SU(N)$ gluodynamics, see  section III of ref.\cite{Fugleberg:1998kk} where connection with approach \cite{Witten:1998uka} is discussed.

As we will see below, instantons are not well-defined objects in this phase.
Indeed, this would contradict $\theta/N$ dependence since each instanton comes
with an integer multiple of $\theta$.
In the holographic model this is resolved by identifying instantons with euclidean
D0 branes wrapping around the $x^4$ direction which tend to shrink
to zero size and disappear \cite{Bergman:2006xn}.

At finite temperature the model exhibits confinement/deconfinement and
chiral phase transitions \cite{Aharony:2006da,Parnachev:2006dn}.
Two possible metrics with euclidean time $t_E$ compactified on a circle with
circumference $\beta$ are (\ref{metric}) and its double analytic continuation,
\beq
ds^2=\left(\frac{U}{R}\right)^\frac{3}{2} \left( (dx_i)^2 + f_T(U)dt_E^2+(dx^4)^2\right)+
       \left(\frac{U}{R}\right)^{-\frac{3}{2}} \left( \frac{dU^2}{f_T(U)}+U^2 d\Omega_4^2\right)
\label{metrict}
\eeq 
where $f_T=1-U_T^3/U^3$ and $\beta= \frac{4\pi}{3}\left(\frac{R^3}{U_K}\right)^\frac{1}{2}$.
Since the two metrics are the same, the comparison of the free energies is simple:
as soon as $U_T>U_K$ the black hole metric (\ref{metrict}) becomes preferred.
This corresponds to confinement/deconfinement transition at $T=1/2\pi R_4$.
The Polyakov loop, which is the order parameter for confinement, vanishes in
the confining phase (\ref{metric}) and has a non-vanishing value in the deconfining
phase (\ref{metrict}).
In the deconfining phase the $x^4$ circle does not shrink to zero
size and Stokes theorem makes it possible to have vanishing $F_2$, which minimizes
the energy  \cite{Bergman:2006xn}. 
That is, in this phase it is possible to have
\beq
   C_1=\frac{\theta}{2\pi} dx^4
\label{dectheta}
\eeq
This leads to $\chi_g=0$ to order $N^0$; this is also consistent with the
fact that instantons are well defined objects in this phase, 
and come with the factor of $e^{i n \theta}$.
In the holographic model this is again a consequence of the
topology in the deconfined phase, where the D0 brane wrapping the $x^4$ circle 
cannot shrink to zero size and disappear.
The factor of $e^{i n \theta}$ in the D0 brane action follows from (\ref{dectheta}).
Hence, we observe that the $\theta$ dependence is different in the confining  and deconfining
phases.
We will also see that such change in the behavior is also supported by analyzing instantons 
in field theory, see next sections.

Another comment we would like to make is the existence of the phase where the glue
is deconfined, but chiral symmetry is broken.
While it is not necessarily true that such a phase exists in QCD (after all, the holographic
model contains two variable parameters, as opposed to $\Lambda_{QCD}$), 
we discuss a field theoretic model with this property to  illuminate the topological charges of the relevant constituents 
in the confining (section V) and the deconfining phases (section IV).

 \section{Large $N$ QCD  at  $T> T_c$  }
 In this section we estimate the value of $T_c$ where the instanton expansion breaks
down and the $\theta$ dependence presumably experiences a sharp change. 
 In the regime $T> T_c$ the $\theta$ dependence 
 is determined by the dilute instanton gas approximation 
(See below for the discussion of the assumptions that are necessary for
this statement to hold).
Applicability  of the instanton expansion (small density)  implies that 
the  $\theta$ parameter enters  partition function in a very simple way $\sim \exp{(-i\theta)}$. 
\exclude{As we shall see in this section  the coefficient $\gamma$ can be always  expressed in terms of temperature and $\Lambda_{QCD}$ as follows,
 \beq
 \label{gamma}
 \gamma\equiv \ln \left(\frac{T}{c\Lambda_{QCD}}\right)
 \eeq
  with calculable coefficient $c\sim 1$ of order one, see below. }
The instanton contribution to the $\eta'$ mass  $\sim \exp{(-\gamma(T) N)}$ 
is exponentially suppressed  for any small (but finite) positive $\gamma>0$.
In contrast: at arbitrary small and negative 
  $\gamma<0$ the instanton expansion obviously breaks down.
The  $\theta $ behavior
presumably drastically changes at $T_c$ determined by 
  \beq
 \label{T_c}
 \gamma (T=T_c) =0 ~~~~~  \Longrightarrow  ~~~~~T_c= c\Lambda_{QCD}.
 \eeq

 \subsection{ Instantons at $T>T_c$  with $\la\bar{\psi}\psi\ra\neq 0$}
 In the following  we will be interested in the instanton density in 
the dilute gas regime at $T> T_c$.
 We assume  that the non-vanishing chiral condensate $\la\bar{\psi}\psi\ra\neq 0$ 
 exists in this region.
  
 The instanton-induced effective action for $N_f$ massless fermions can be easily constructed. In particular, 
  for $N_f=2$ flavors, $u, d$ the corresponding expression takes the following form, 
\cite{tHooft,SVZ,Vainshtein:1981wh,Gross:1980br},
\begin{eqnarray}
 \label{inst_vertex}
   L_{\rm inst}&=&e^{-i\theta} \int\!d\rho\, n(\rho) 
  \biggl(\frac{4}{3}\pi^2\rho^3\biggr)^2 \biggl\{
  (\bar u_R u_L)(\bar d_R d_L) + \\
  &+& {3\over32} \biggl[ (\bar u_R\lambda^a u_L)(\bar d_R\lambda^a d_L)
  - {3\over4}(\bar u_R\sigma_{\mu\nu}\lambda^a u_L)
    (\bar d_R\sigma_{\mu\nu}\lambda^a d_L) \biggr]
  \biggr\}  + {\rm H.c.}\nonumber
\end{eqnarray}
We wish to study this problem at nonzero temperature and small chemical
potential $\mu$ (to be discussed later in the text) for $T> T_c$, and we use the standard formula
for the instanton density at two-loop order \cite{shuryak_rev}
\begin{eqnarray}
\label{instanton}
 n(\rho)= C_N(\beta_I(\rho))^{2N} \rho^{-5}
 \exp[-\beta_{II}(\rho)]  
 \times  \exp[-(N_f \mu^2 + \frac13 (2N+N_f) \pi^2
 T^2)\rho^2], 
\end{eqnarray}
where
\begin{eqnarray}
  C_N = \frac{0.466 e^{-1.679N} 1.34^{N_f}}{(N-1)!(N-2)! },~~
\beta_I(\rho)&=&-b \log(\rho\Lambda_{QCD}), ~~
\beta_{II}(\rho)=\beta_I(\rho)+\frac{b'}{2b} \log\left(\frac{2
  \beta_I(\rho)}{b}\right),  \nn \\ 
b&=& \frac{11}3 N-\frac23 N_f,~~~~~
 b'=\frac{34}3 N^2-\frac{13}3 N_f
N +\frac{N_f}{N}. \nn
\end{eqnarray}
This formula  contains, of course, the standard instanton classical action $\exp(-8\pi^2/ g^2(\rho))
\sim  \exp[-\beta_{I}(\rho)]  $
which however is hidden as it is   expressed in terms of $\Lambda_{QCD}$
rather than in terms of coupling constant $g^2(\rho)$.
By taking the average of eq.(\ref{inst_vertex}) over the
 state  with nonzero vacuum expectation value for the chiral 
 condensate $\la\bar{\psi}\psi\ra\neq 0$,  one finds  the following expression
 for the instanton induced interaction, defined as  $V_{\rm inst}(\varphi)\equiv  -\la L_{\rm inst}(\varphi)\ra$,
\bea
\label{Vinst2}
  V_{\rm inst}(\varphi)=- \biggl[\la\bar{\psi}\psi\ra^{N_f}\cos(\varphi-\theta)\biggr]\cdot  \int\!d\rho\, n(\rho)
  \biggl({4\over3}\pi^2\rho^3\biggr)^{N_f}  \biggl(\frac{1}{2}\biggr)^{N_f-1} 
 = -a\cdot \Lambda_{QCD}^4\cos(\varphi-\theta),
\eea 
where we introduced small dimensionless parameter $a\ll 1$ which   essentially 
governs the relevant physics. We 
  assumed factorization  for the chiral condensates in large $N $ limit in deriving (\ref{Vinst2}).
Hence the square bracket in the eq. (\ref{inst_vertex}) vanishes.
 We also assumed that the condensates for all flavors are equal,
   $\la\bar u_R u_L\ra=\la\bar d_R d_L\ra=...= \frac{1}{2}\la\bar{\psi}\psi\ra$.      
   Finally we introduced a singlet phase of the chiral condensate     $\varphi(x)$ which we iddentify
    with the $\eta'(x)$ field in the standard way, $\bar{\psi}_R\psi_L\sim e^{i\varphi(x)/N_f}$.    
    As expected, the $\eta'$ field $\varphi(x)$ enters the lagrangian in unique combination with
    $\theta$ as $(\varphi-\theta)$ which is consequence of the anomalous Ward Identity\footnote{   In the presence of the massless chiral fermions the $\theta$ dependence as is known goes away in full QCD.  To avoid any confusions later in the text   we remark here that  our  discussions  of $\theta$ dependence in this paper   deals exclusively with the dynamics of gluons when  light 
   fermion degrees of freedom are  frozen such that essentially
 we analyze the  $\theta$ dependence in gluodynamics rather than in full QCD. 
 In different words, we assume a quenched  approximation for $N_f\ll N$. 
 Precisely the $\theta $ dependence    in quenched  approximation plays a crucial role  in  understanding of the  dynamics of   strongly interacting systems.}.
   
The mass of the $\varphi$ field in the chiral limit is determined by the instanton density in this phase and  is  expressed in terms of parameter $a$, 
\beq
  L = -\frac{1}{2}f_{\eta'}^2 (\vec{\nabla}\varphi)^2  
   + m_{\eta'}^2f_{\eta'}^2\cos(\varphi-\theta), ~~~~~ 
   m_{\eta'}^2f_{\eta'}^2\equiv a\cdot \Lambda_{QCD}^4, 
  \label{cos}
\eeq
where   $f_{\eta'}$   is defined in the standard way
as a normalization of $\varphi$ field, $\eta' (x)\equiv f_{\eta'}\varphi (x)$, and in general
$f_{\eta'}(T)$ depends on temperature $T$ (though it will not be explicitly shown later in the text).
We also keep only the lowest Matsubara frequency for $\varphi$ field in the environment 
with $T\neq 0$ 
which ensures the validity of the static  approximation   for all interactions involving $\varphi$.
 This effective lagrangian, is by definition a Wilson type lagrangian for the light   $\eta'$ field
 which is valid  as long as $\eta' $ field is light, 
 \begin{eqnarray}
  \label{critical}
m_{\eta'}\sim \sqrt{a}\Lambda_{QCD} \ll  \cdot \Lambda_{QCD},
\end{eqnarray}
   
In the large $N$ limit parameter $a\sim e^{-\gamma N} $ is exponentially suppressed\footnote{See also \cite{Kharzeev:1998kz} for   earlier discussions  on the subject.}
for  temperatures
above $T_c$, $a\ll 1$ and 
the instanton expansion converges.
For $T<T_c$  the instanton expansion makes no sense (breaks down) and the expansion parameter becomes large $a\gg 1$.
We assume that  $\theta $ dependence sharply changes at $T=T_c$.  
We estimate the value of  $T_c$   by equalizing $\gamma=0$ according to eq.
(\ref{T_c}) see below.
\footnote{It is conceivable that the
phase transition and   sudden change in $\theta$ behavior occur at the same point $T_c$ for any finite $N$,
and not only for  $N=\infty$.
This assumption  allows us to make some reasonable estimate for  $T_c$ for  finite $N$.
By obvious reasons, an estimate  of $T_c$ at  finite $N$  suffers  from some  inherent  uncertainties. Indeed, $T_c$ in this case  is determined  by an approximate  condition $a\sim 1$ in contrast with precise equation (\ref{T_c}) valid
for $N=\infty$ case. The  condition $a\ll 1$ implies that the 
$\eta'$  field is much 
lighter than   all  other degrees of freedom in the system in the chiral limit and condition (\ref{critical})
is satisfied.
It is clear that this condition can be always satisfied for sufficiently large $N$ where parameter 
$a$ is exponentially small at $T>T_c$.  When $T$ becomes close to $T_c$ from above, parameter $a$
increases and becomes order of unity at some point.
 This is precisely the region where instanton approximation 
breaks down. Therefore, according to our logic, the $\theta$  dependence 
may sharply  change  here. We identify this point    where $a\sim 1$   with the point of the 
  phase transition   $T_c$.   
Of course we do not know the precise coefficient here (magnitude  of  $a$ could be, for example 3, instead of 1), but the extracting of a large power 
in  such an estimate , $T_c\sim \Lambda_{QCD}\cdot  a^{-\frac{3}{11N}}$ should not produce
a large error for estimation of $T_c$ even for physically relevant case $N=3$. }
In deriving the low energy effective lagrangian for the $\eta'$ field we should, 
in principle, use the exact formula for the instanton density and not (\ref{instanton})
which is only valid in the two-loop approximation.
\exclude{
This is not a problem as long as $T\gg T_c$, where $n(\rho)$ is strongly peaked near $\rho\sim 1/T\gg \Lambda_{QCD}$,
and the euclidean action has the instanton solution as its sharply defined minimum.
As the temperature is lowered, perturbative expansion around the instanton field
configuration becomes unreliable.
It is possible that $n(\rho)$ ceases to be integrable at some value of temperature
of order $\Lambda_{QCD}$.
We assume that a different option is realized,  $n(\rho)$ remains integrable for
temperatures above $T_c$ determined by (\ref{T_c}), and the resulting function
$\gamma(T)$ is a monotonic function of $T$ which crosses zero at $T=T_c$.}
We assume that the perturbative corrections for $T\sim T_c$, although large,
do not drastically change the physics.
Then we will see that for any $T>T_c$ the dilute instanton approximation is valid,
since the average distance between the instantons is parametrically larger
then their size, see eq.(\ref{T}) below. 
To reiterate, we do not know how to do an honest instanton calculation in
the close vicinity  of $T_c$, but we assume that the perturbative expansion around 
the instanton field configuration can still be performed and would yield $a\sim e^{-\gamma(T) N}$
where $\gamma(T)$ is a monotonic function vanishing at $T=T_c$.
Then, for $T>T_c$ the dilute instanton gas approximation is good, for $T<T_c$
it is no longer valid, while $T=T_c$ describes the phase transition point with drastic changes 
in $\theta$ behavior.

We should also note that one can  estimate  $T_c(\mu)$
for  non zero chemical potential $\mu\neq 0$ 
as long as the chiral condensate
\exclude{\footnote{ Even if chiral symmetry is not broken at $T>T_c$
the $U_A(1)$ symmetry is always explicitly broken due to the instantons. It implies that
the product $ \la(\bar{\psi}\psi)^{N_f}\ra $ does not vanish even if the chiral condensate for each flavor vanishes. Therefore, our arguments still hold in the case when the chiral symmetry is not 
spontaneously broken at $T>T_c$.}}
does not drastically varies with $\mu$,
which we assume to be the case at least for sufficiently small $\mu$. It allows
us to estimate not only a single point $T_c$ on the phase diagram but entire phase transition line $T_c(\mu)$ for sufficiently small $\mu\ll T_c$.

\subsection{Numerical estimates} 
  First, we estimate  the critical temperature $T_c$ 
  by solving eq. (\ref{T_c}) and calculating coefficient $c$ using the expression for the instanton density (\ref{instanton}). As the first approximation (which greatly simplifies computations) we neglect all
  $\log (\rho\Lambda_{QCD})$ factors in evaluating $\int d\rho$ integral. In this case the integral can be 
  computed analytically and  the limit $N\rightarrow\infty$ can be easily evaluated. The result for  the  instanton contribution   takes  the following form (as expected)
  \beq
  \label{gamma_N}
 V_{\rm inst}(\varphi)\sim e^{-\gamma N} \cos(\varphi-\theta),~~~~ \gamma=\Bigl[\frac{11}{3}
 \ln \left(\frac{\pi T}{\Lambda_{QCD}}\right)-1.86\Bigr],
  \eeq
  where we neglected all powers $N^p$ in front of $ e^{-\gamma N}$ (as it does not have any impact on computation of $T_c$ at $N=\infty$)  and used the standard Stirling formula 
    \beq
  \label{stirling}
  \Gamma (N+1)=\sqrt{2\pi N}N^N e^{-N}\left(1+\frac{1}{12N} +O(\frac{1}{N^2}\right)
  \eeq
  to evaluate $N\rightarrow\infty$ limit.
  
  There are three  main reasons for a generic structure
(\ref{gamma_N})  to emerge. First of all, 
there is an   exponentially large ``$T-$ independent"  contribution, expressed as 
$e^{+1.86 N}$ in eq. (\ref{gamma_N}). This term 
   basically describes  the entropy of the configuration such as  
 number of embedding $SU(2)$ into $SU(N)$ etc. Secondly, 
 there is  a ``$T-$ dependent" contribution to  $V_{\rm inst}(\varphi)  $ which comes  from  $\int n(\rho) d\rho$ integration (\ref{instanton}).  It   is proportional to
  \beq 
  \label{rho}
     \left(\frac{\Lambda_{QCD}}{\pi T}\right)^{\frac{11}{3}N}=\exp\Bigl[-\frac{11}{3}N
\cdot \ln \left(\frac{\pi T}{\Lambda_{QCD}}\right)\Bigr].
 \eeq
Finally,  all fermion related contributions such as a  chiral condensate or non-vanishing mass term    enter the instanton density as follows $ \sim\la\bar{\psi}\psi\ra^{N_f}\sim e^{N\cdot \left(\kappa\ln  |\la\bar{\psi}\psi\ra|\right)} $. For $\kappa\equiv\frac{N_f}{N}\rightarrow 0$ this term  obviously leads to a sub leading effects $1/N$ in comparison
  with two  main terms in the exponent (\ref{gamma_N}). Therefore, such terms can be neglected as they do not change any estimates
  at $N=\infty$. It is  in accordance with the general arguments suggesting that the fundamental fermions can not change the dynamics of the relevant gluon configurations  as long as $N_f\ll N$. Indeed, the formula for  $\gamma(T)$ for pure gluodynamics is given by the same expression (\ref{gamma_N})     as it should be for $N_f\ll N$.
  If the chiral condensate vanishes, one can replace it by  a small but  nonzero value for  the quark's mass to proceed with our calculations. It would not alter the   equation (\ref{gamma_N}). 
  Therefore our estimate below (\ref{T_c_N}) is not  effected whether the chiral condensate develops or   not. In different words, we essentially study a pure gluodynamics. Our treatment of the problem is equivalent of    a quenched  approximation for $N_f\ll N$.    
  The fermion fields in the present study play an auxiliary (not a dynamical) role  as a  probe of the  topological charges relevant for the phase transition as will be discussed in the next sections. When a number of fermions increases and $N_f\sim N$ we can not proceed with the estimations as we have done above.
In this case we do anticipate a strong dependence on fermion properties such as quark's masses and the chiral condensate as argued in recent paper \cite{Zhitnitsky:2008ha}. Lattice simulations also suggest that for physical values for the quark's masses one should expect  a smooth crossover rather than a first order phase transition, see e.g. \cite{Vicari:2008jw,Kogut:2004su}. 
We remind the reader that   for pure gluodynamics which is the main subject of the present paper for all $N\geq 3$ and small number of flavors $N_f\ll N$ one should expect the first order phase transition.

  From discussions above it should be obvious that 
  there will be always a point $T_c$ where  two leading contributions with  exponential $e^N$ dependence   cancel each other.
As a result,  at $N\rightarrow \infty$ for $T>T_c$ the instanton gas is dilute with density
$e^{-\gamma N}, ~\gamma>0$ which ensures  a  nice $\cos\theta$ dependence,   while  for $T<T_c$ 
the expansion breaks down, and $\theta$ dependence must sharply change at $T<T_c$.  We  identify such sharp changes with first order phase transition.

  As explained above, the critical temperature is determined by condition $\gamma =0$.
Numerically, at one loop level approximation,
  it happens at 
   \beq
  \label{T_c_N}
  \gamma=\Bigl[\frac{11}{3}  \ln \left(\frac{\pi T_c}{\Lambda_{QCD}}\right)-1.86\Bigr]=0
 ~~~  \Rightarrow ~~~T_c (N=\infty)\simeq 0.53 \Lambda_{QCD},
  \eeq
  where $ \Lambda_{QCD} $ is defined in the Pauli -Villars scheme. 
A few remarks are in order. \\
{\bf a.} Our computations are carried out in the regime where the instanton 
density $\sim \exp(-\gamma N) $ is parametrically suppressed at $N=\infty$.
From eq. (\ref{gamma_N}) one can obtain the following expression for 
instanton density in vicinity of $T>T_c$, 
   \beq
   \label{T}
   a \sim \cos(\varphi-\theta) \cdot e^{-\alpha N \left(\frac{T-T_c}{T_c}\right)}, ~~~~ 1\gg \left(\frac{T-T_c}{T_c}\right)\gg 1/N.
   \eeq
where $\alpha$ is a numerical coefficient of order one.
 Such a behavior does  imply that the dilute gas approximation is justified even in close vicinity of $T_c$ as long as $\frac{T-T_c}{T_c}\gg \frac{1}{N}$.    In this case the diluteness parameter  remains small\footnote{This should be  contrasted with the  the standard requirement for
    finite $N$ when the 
   condition  $a\sim (\Lambda_{QCD}/T)^b \ll 1$ can be only   achieved when the temperature is very large,  $T\gg   \Lambda_{QCD}$. 
   For large $N$ the condition $a\ll 1$  is satisfied as long as $\frac{T-T_c}{T_c}\gg \frac{1}{N}$ as can be seen from eq.(\ref{T}).}
 even in the close vicinity of $T_c$.   Therefore, the $\theta$ dependence, which is sensitive to the 
 topological fluctuations only,
  remains unaffected all the way down to the temperatures very 
    close to the phase transition point, $T=T_c+ O(1/N)$. We can not rule out, of course, the possibility that the  perturbative 
   corrections may change our numerical estimate for $T_c$. 
However, we   expect that a qualitative picture of the phase transition advocated in this 
paper remains unaffected  as a result of these    corrections   in dilute gas regime.\\
{\bf b.} In our estimate for $T_c$   we neglected $(\log \rho\Lambda_{QCD})^k$ 
in evaluating of the $\int d\rho  $ integral. One can easily take into account the corresponding
contribution by notice that $\int d\rho  $ is saturating at $\rho\simeq (\pi T)^{-1}$. The corresponding 
correction changes  our estimate (\ref{T_c_N}) very slightly, and it will be ignored in what follows.  Numerical  smallness of correction is   due to the strong  cancellation
between the second loop contribution in the exponent (term proportional to $b'/ b$) and the first loop 
contribution in the pre-exponent in eq. (\ref{instanton}).  \\
{\bf c.}  The transition to a different scheme leads to very large changes in the instanton density. For example, transition to the so-called MS -scheme is achieved by replacing 
$e^{-1.679N}$ in the expression for $C_N$, see eq. (\ref{instanton}), as follows  $e^{-1.679N}
\rightarrow e^{(-1.679 + 3.721)N}$ with  a number of other changes, see e.g.\cite{Vainshtein:1981wh}. The corresponding  results would be expressed in terms of $ \Lambda_{QCD}^{MS}$, where MS stands for  MS -scheme,
to be distinguished from $ \Lambda_{QCD} $ which   is defined in the Pauli -Villars scheme
and will be used through this  paper.
We shall not elaborate on these numerical issues  in the present work.\\
{\bf d.} Unfortunately, we can not compare our calculations with the precise lattice results
\cite{Lucini:2003zr} for the ratio $T_c/\sqrt{\sigma}$ at large $N$ as we compute  $T_c$ in de-confined phase  where the string tension $\sigma$ vanishes.\\
{\bf e.} As expected, the result (\ref{T_c_N}) does not depend on a number of flavors $N_f$
nor does it depend on the magnitude of 
the chiral condensate in $N=\infty $ limit as our treatment of the problem corresponds essentially pure YM  computations.\\
{\bf f.} For finite but large 
$N\gg N_f$ the corresponding numerical estimates for $T_c$  can also be given. It can be estimated 
from condition  $a\sim 1$.  However, 
  numerical estimates  in this case would   depend on the value of the $U_A(1)$ condensate $ a\sim\la(\bar{\psi}\psi)^{N_f}\ra $ which is not well-known for $T>T_c$.  
Therefore, 
  we shall not discuss the corresponding  numerical estimates in the present work.\\
   {\bf g.}  A similar procedure for   estimation
of the critical  chemical potential $\mu_c$ for confinement -de-confinement phase transition at finite 
$N, N_f$ at $T\sim 0$     has been previously used in ref.\cite {TZ} where the analogous  arguments on drastic changes   of $\theta$ at $\mu=\mu_c$ have been presented, see also a review paper\cite{Zhitnitsky:2006sr}.\\
   {\bf h.}  
   Once $T_c$ is fixed    one can compute the entire line of the phase transition $T_c(\mu)$  for    relatively small $\mu \ll T_c$ for large but 
finite $N\gg N_f$.
  Indeed, in the weak coupling regime at $T>T_c$  the $\mu$ dependence of the instanton density is determined
by a simple insertion   $\sim \exp[- N_f \mu^2\rho^2]$ in the expression
for the density (\ref{instanton}). In the leading loop order  $T_c(\mu)$ varies as follows,
\beq
\label{mu}
T_c(\mu)=T_c(\mu=0)\Bigl[1- \frac{3N_f\mu^2}{4N \pi^2 T_c^2(\mu=0)} \Bigr], ~~ \mu\ll \pi T_c, ~~~ N_f\ll N.
\eeq
As expected, $\mu$ dependence goes away in large $N$ limit in agreement with general large $N$ arguments\cite{Toublan:2005rq}. This formula is in excellent agreement  
with numerical computations  \cite{de Forcrand:2002ci,de Forcrand:2003hx,mu} which  show very  little changes of the critical temperature $T_c$ with $\mu$ for sufficiently small chemical potential.
In particular, even for the case $N_f= 2, ~N=3$ where the expression (\ref{mu}) is not expected  to give a good numerical
estimate, it still works amazingly  well even for $N=3$. Indeed, 
 the result quoted in  \cite{de Forcrand:2002ci}  can be written as 
 $$T_c(\mu)^{lat}=T_c(\mu=0)^{lat}\Bigl[1- 0.500 (38) \frac{\mu^2}{  \pi^2 T_c^2(\mu=0)^{lat}} \Bigr], ~~~ N_f=2, ~~~N=3.$$
It should  be compared with our theoretical prediction (\ref{mu}) for this case
  $$T_c(\mu)^{th}=T_c(\mu=0)^{th}\Bigl[1- \frac{1}{2} \frac{\mu^2}{  \pi^2 T_c^2(\mu=0)^{th}} \Bigr].$$
    {\bf i.} It is naturally to expect
that the phase transition line $T_c(\mu)$ at $\mu\ll T_c$ from (\ref{mu})
connects with  the phase transition line at very large $\mu_c\sim \sqrt{N}$
as estimated in a recent paper \cite{Zhitnitsky:2008ha}, 
\bea
\label{T1}
\mu_c(T)\simeq \mu_c(T=0)\Bigl[1- \frac{N \pi^2T^2}{3N_f \mu_c^2(T=0)} \Bigr], ~~~~
 \sqrt{N}T\ll \mu_c, ~~~ N_f\ll N  ,
 \eea
 where
$\mu_c (T=0)  \simeq 1.4\cdot \Lambda_{QCD} \sqrt{\frac{N }{N_f}}$ at   $ N_f\ll N$  \cite{Zhitnitsky:2008ha}.
 This expectation is motivated by  the observation  that the nature  for the phase transition along the entire line is one and the same: it is drastic changes of $\theta$ dependence when the phase transition line is crossed. Therefore, we believe that the entire   line is the first order
 phase transition as long as $N_f\ll N$.

 \section{Dual representation }
 The main goal of this section is to present the low energy effective lagrangian for $\eta'$ field
 (\ref{cos}) in the dual form. The $\eta'$ field will play a crucial role in the following two sections. 
 As we shall see in a moment the $\eta'$-field  is a perfect  probe of the glue configurations. This field will  help us to investigate the topological charges of the constituents in both phases, and therefore it will  help us to   interpret the nature of the
 phase transition whose critical temperature $T_c$  was computed in the previous section. 
In section II we discussed a holographic model with nonvanishing chiral condensate.
Here we consider a field theoretic model with this property.

 \subsection{Coulomb Gas Representation: formal derivation }

The effective low energy dense-QCD Lagrangian (\ref{cos})
is the sine-Gordon (SG) Lagrangian.
Many of the special properties of the SG theory apply.
One of these properties is the   admittance of a
Coulomb gas (CG) representation for the partition function.  
   Although this is a four-dimensional theory  at nonzero temperature $T$ (rather than two dimensional,
 where all known exact results regarding SG model were discussed) and questions
 about renormalizability of the theory may come to mind, there are no
 such issues here since the effective action is a low energy
 one. Following the usual procedure for mapping a statistical CG model
 into the field theoretic SG model, the CG picture that arises from the
 effective low energy $QCD$ action, Eq. (\ref{cos}), will be derived in this
 section. The statistical model contains
 some  charges which appear due to the presence of  cosine
 interaction in the field theory model. The physical meaning of these
 charges will be illuminated in the next section by analyzing the corresponding measure
 of the statistical ensemble.

The mapping between the SG theory and its CG representation is
well known. All we need to do is to reverse the derivation
of SG functional representation of the CG in Ref.\ \cite{Po77}.
The partition function corresponding to
the Lagrangian (\ref{cos}) is given by\footnote{To be precise,
the path integral in Eq.\ (\ref{Z}) should be understood as an
integral over {\em low-momentum} modes of $\varphi$ only.  The upper
limit of the momentum of $\varphi$ is the ultraviolet cutoff of
the effective Lagrangian (\ref{cos}), which should be taken as some
scale smaller than $T$.  Only tree graphs contribute to $Z$ so
there is no dependence on the precise value of the cutoff.}
\beq
\label{Z}
Z = \int \D \varphi\,  e^{-\int d^3x\,\int^{\beta}_0d\tau\,L_E} 
=  \int \D \varphi \ 
e^{ - \frac{1}{2T}f_{\eta'}^2 \int d^3x (\vec{\nabla }\varphi)^2}\,
e^{ \lambda \int d^3x \cos\left(\varphi(x)-\theta\right)} \, ,
\eeq
where we introduced fugacity for the CG ensemble to be defined as, 
\beq
\label{lambda}
\lambda \equiv \left(\frac {\Lambda_{QCD}}{T}  \right)a\Lambda_{QCD}^3
\eeq
$L_E$ is the Euclidean space Lagrangian.  Leaving alone the integration over $\varphi(x)$ for a moment,
we expand the last exponent in Eq.\ (\ref{Z}), represent the
cosine as a sum of two exponents and perform the binomial expansion:
\bea
\label{seriesexp}
{\rm e}^{ \lambda \int d^3x \cos(\varphi(x)-\theta)}
&=&
\sum_{M =0}^\infty \frac{(\lambda/2)^M}{M!} 
\left( \int d^3x\sum_{Q =\pm1}\,e^{iQ(\varphi(x)-\theta)} 
\right)^M \nn \\
&=&
\sum_{M_\pm=0}^\infty \frac{(\lambda/2)^M}{M_+!M_-!}  
\int d^3x_1 \ldots \int d^3x_M\ 
e^{i\sum_{a=0}^M Q_a(\varphi(x_a)-\theta)}\ .
\eea
The last sum is over all possible sets of $M_+$ positive and $M_-$ 
negative charges $Q_a=\pm1$. The last line in Eq.\ (\ref{seriesexp})
is a classical partition function of an ideal gas of $M=M_++M_-$ identical 
(except for charge) particles of charges $+1$ or $-1$ 
placed in an external potential given by $i(\theta-\varphi(x))$. 
It is easy to see
that (for a constant or slowly varying potential) the average 
number of these particle per unit of 3-volume $\langle M \rangle/V_3$, 
i.e., the density, is equal to $\lambda$. 
Thus making $\lambda$ small one can make the gas
arbitrarily dilute, which is precisely the physical meaning of the fugacity.
From this observation, one can immediately see that the average distance
between charges $Q_a$ is  $ \lambda^{-1/3} $.

While $\theta$ can indeed be viewed as an external potential for the gas
(\ref{seriesexp}), $\varphi(x)$ is a dynamical variable,
since it fluctuates as signified by the path integration in (\ref{Z}).
For each term in (\ref{seriesexp}) the path integral is Gaussian and
can be easily taken:
\beq
\label{intphi}
\int \D \varphi\, e^{ - \frac{1}{2T}f_{\eta'}^2 \int d^3x (\vec{\nabla} \varphi)^2}\,
e^{i\sum_{a=0}^M Q_a(\varphi(x_a)-\theta)}
=
e^{-i\theta\sum_{a=0}^M Q_a}\,
e^{-{T\over f_{\eta'}^2}\sum_{a>b=0}^M  Q_a  G(x_a-x_b) Q_b
}\ .
\eeq
We see that, for a given configuration of charges $Q_a$, $-i\varphi(x)$
is the Coulomb potential created by such distribution.\footnote{
One notices that the term $a=b$ in the double sum (\ref{intphi}) is dropped.
This is the self-interaction of each charge. It would renormalize the
fugacity $\lambda$ by a factor $\exp(-G(0)/( f_{\eta'}^2))$. This
factor should be dropped as it represents contribution of very
short wavelength fluctuations of $\varphi$.
Such fluctuations have to be cutoff at the scale $1/T$.
The self-energy of the
charges comes from a much smaller scales  which are
already calculated and contained in $a$.
}
The function $G(x)$ is the solution of the three-dimensional Poisson
equation with a point source (the inverse of $-\vec{\nabla}^2$):
\beq
\label{G}
G(x_a - x_b) = {1\over 4\pi |\vec{x}_a-\vec{x}_b|} \, .
\eeq
Thus we obtain the dual CG representation for the partition function
(\ref{Z}):
\beq
\label{CG}
Z = \sum_{M_\pm=0}^\infty \frac{(\lambda/2)^M}{M_+!M_-!}  
\int d^3x_1 \ldots \int d^3x_M\ 
e^{-i\theta\sum_{a=0}^M Q_a}\,
e^{-{T\over  f_{\eta'}^2}\sum_{a>b=0}^M  Q_a  G(x_a-x_b) Q_b
}\ .
\eeq
The two representations of the partition function (\ref{Z}) and
(\ref{CG}) are equivalent.

 Note that the 
 physical meaning of $\lambda$ is the 
 the fugacity of the system with charges   $Q $  
  and it is proportional to  
 $a$ which is small in the regime under discussion.
 There are several important features of the action (\ref{CG})
 which should be noted. Firstly,  the 
 total $Q$ charge of the configuration, $Q_T$ appears together with
 $\theta$  which we kept as a free parameter, see  (\ref{CG}).
 Such a dependence will play an important role in the following identification
 of $Q$ charges as the topological charges, see below. 
  The    $\theta$-dependence in CG representation (\ref{CG})
gives an overall phase factor for each configuration.
 Finally, our dimensional parameters $\lambda$ and $\vec{x}_M$ 
 come into the expression (\ref{CG}) in the combination $\lambda d^3x$ 
 which is  nothing but the coefficient in front of the instanton
 contribution to the effective action   $S_{\rm inst}= \int^{\beta}_0d\tau\int d^3x  V_{\rm inst}(\varphi) =-
 \int d^3x\lambda \cos(\varphi-\theta)$ at nonzero temperature $T$, see eq. (\ref{Vinst2}) and 
 definition of $\lambda$ (\ref{lambda}).

\subsection{Physical Interpretation}

The charges $Q_a$ were originally introduced in a rather  formal manner
so that the QCD effective low energy Lagrangian can be written in the
dual CG form (\ref{CG}). However, now the
physical interpretation of these charges becomes clear: since
$Q_{\rm net}\equiv \sum_a Q_a$ is the total charge and it appears in
the action
multiplied by  $\theta$ [see Eq.\ (\ref{CG})], one
concludes that $Q_{\rm net}$ {\it is} the total topological
charge  of a given
configuration.  Indeed, in QCD the $\theta$ parameter appears in the
Lagrangian only in the combination with the topological
charge density $-i\theta G_{\mu\nu}
  {\widetilde G}_{\mu\nu}/(32\pi^2)$.  
It is also quite obvious that  each charge 
$Q_a$ in a 
given configuration should be identified with an
integer topological charge well localized at the point $x_a$. This, 
by definition,  corresponds to a small instanton positioned at $x_a$ (to be precise, ``caloron" at 
temperature $T\neq 0$ which has topological charge $Q=1$ and action $8\pi^2/g^2$ independent
of temperature,  see  \cite{Gross:1980br,shuryak_rev} for review).
To support this identification we note that every particle with
charge $Q_a$ brings along a factor of fugacity $\lambda\sim a$ (\ref{lambda})
which contains the classical one-instanton suppression factor
$\exp(-8\pi^2/ g^2(\rho))$ in the density of instantons (\ref{Vinst2})
if one restores the instanton density in terms of coupling constant $\exp(-8\pi^2/ g^2(\rho))$ rather than 
directly in terms of $\Lambda_{QCD}$ which is used in eq.  (\ref{instanton})
and which is more convenient for numerical estimates. 
  
This identification is also supported by the following observation:
every extra  particle with charge $Q_a$ brings an additional
weight $e^{-i\theta Q_a}$ to the partition function.
This is certainly the most distinguishable feature of the
non-zero  topological charge configuration.

The following hierarchy of scales exists in such an instanton ensemble
for temperatures slightly higher than $T_c$.
The typical size of the instantons $\rho \sim T^{-1}\sim \Lambda_{QCD}^{-1}$
The average distance between the instantons $\bar{r}=\lambda^{-1/3}=
\Lambda_{QCD}^{-1}a^{-1/3}$ is much larger than both the average size 
of the instantons and the cutoff $T^{-1}$.
The largest scale is
the Debye screening length in the Coulomb gas,
$r_D =\Lambda_{QCD}^{-1} a^{-1/2}$.
This coincides with the static correlation length of the $\varphi$ field, 
which is precisely $\eta'$ mass. It is important that the Debye screening length $r_{D}$
is parametrically larger than the average distance between the instantons $\bar{r}$,
therefore large number of instantons can be accommodated within  the volume determined by 
the Debye screening length $r_{D}$ which justifies our Coulomb gas interpretation, at least in large 
$N$ limit.
In short:
\beq
\begin{array}{ccccccc}
\label{hierarchy}
\mbox{(size, $\rho$)}& \ll& \mbox{(distance, $\bar{r}$)}&\ll& \mbox{(Debye, $r_D$)}\\
\frac{1}{T} &\ll &\frac{1}{\Lambda_{QCD}\sqrt[3]{a}}&\ll& \frac{1}{\Lambda_{QCD}\sqrt{a}}
\end{array}
 \eeq
Due to this hierarchy, ensured by small  $a\ll1$, we acquire 
analytical control. In reality, of course, $a\sim  \left({\Lambda_{QCD}\over T}\right)^{N}\sim e^{-N}\ll 1$ is parametrically very small only at very large $N$ while $\left({\Lambda_{QCD}\over T}\right)\leq 1$ could be very close to 1 from below. It implies that at $N=3$ all scales could be numerically very close to each other.

It is also quite interesting that,
although the starting low-energy effective Lagrangian contains only 
a colorless field $\varphi$, we
have ended up with a representation of the partition function in which 
objects carrying color (instantons, their interactions and
distributions) can be studied.
In particular, from the discussion  above, one can immediately
deduce that $II$ and $I\bar I$
interactions are exactly the same up to a sign 
and are Coulomb-like at large distances.
  
This looks highly nontrivial
since it has long been known  that 
at the semiclassical level
an instanton interacts only
with anti-instantons 
but not with 
another instanton carrying a topological charge of the same sign.
As we demonstrated above it is not true any more at the quantum level
in the presense of the $\eta'$ field.
Indeed, what we have found is that the interactions between 
dressed (as opposed to bare) instantons and anti-instantons
after one takes into account
their  classical and quantum interactions,
after integration over their all possible sizes and color orientations, after
accounting for the interaction with the background chiral condensate 
 must become very simple at large distances as 
explicitly described by Eq.\ (\ref{CG}).
It is impressive how the problem which
looks  so complicated in terms of the original bare (anti)instantons, 
becomes so simple in terms of the dressed (anti)instantons
when all integrations over all possible sizes, color orientations
and interactions with background fields are properly accounted for!

Such a simplification of the interactions is
of course due to the presence of almost massless
pseudo-Goldstone boson $\eta'$ which couples to the topological charge.
When the instanton gas becomes very dilute all semiclassical 
interactions (due to zero
modes) cannot contribute much, since they fall off with distance
faster then the Coulomb interaction mediated by $\eta'$. 
On the other
hand, when the instanton density increases  
when $T$ is getting smaller,  the Coulomb interaction becomes more screened and, as
the Debye length becomes comparable to the inter-instanton distances,
we lose analytical control. Based on this picture one can estimate the critical
temperature $T_c$ where this transition must happen. It corresponds to the same condition $a\sim 1$
 discussed  previously in  section III.

We collect here the most important results of the present section  based on CG representation (\ref{CG}):\\
{\bf a. } Since
$Q_{\rm net}\equiv \sum_a Q_a$ is the total charge and it appears in
the action
multiplied be the parameter $\theta$, one
concludes that $Q_{\rm net}$ {\it is} the total topological
charge  of a given
configuration.\\
{\bf b.} Each charge 
$Q_a$ in a 
given configuration should be identified with an
  integer topological charge  $Q_a=\pm 1$ well localized at the point $x_a$. This, 
by definition, 
corresponds to a small instanton (caloron at $T\neq 0$) positioned at $x_a$.\\
{\bf c.} While the starting low-energy effective Lagrangian contains only 
a   colorless field $\varphi$  we
have ended up with a representation of the partition function in which 
  objects carrying color (the instantons)   can be studied.\\
{\bf d.}
In particular, $II$ and $I\bar I$
interactions (at very  large distances) are exactly the same up to a
sign, order $g^0$, 
and are Coulomb-like.  This is in  contrast with semiclassical expressions
when  $II$ interaction is zero and $I\bar I$  interaction is order $1/g^2$.\\
{\bf e.} The very complicated picture of the    bare  $II$ and $I\bar I$
interactions 
becomes very simple for  dressed   instantons/anti-instantons
when all integrations over all  possible sizes, color orientations
and interactions with background fields are properly accounted for.\\
{\bf f.} As expected, the ensemble of small $\rho\sim T^{-1} $ instantons can
not produce confinement because small  instantons can not produce a 
 correlation at arbitrary large distances which is a crucial feature of the confinement. 
 This is in accord with the fact that there is no confinement  in the high temperature phase.\\
 {\bf g.} Physical interpretation of the CG representation (\ref{CG}) is simple. 
 The $\eta' -$ field being a dynamical field couples
 to the topological charge $Q$ exactly as $\theta$ parameter does due to the 
 specific combination $(\varphi (x)-\theta)$ which appears in the low energy  lagrangian.   
 In the dual language the $\eta'$ mass emerges as  a result of Debye screening in the plasma of topologically  charged instantons (interacting via $\eta'$ Coulomb exchange ) similar to the well-known effect of generating the photon's mass
 in the ionized plasma due to the Coulomb interaction of  charged particles. In our case instead of a conventional  vector photon we are dealing with  pseudo scalar $\eta'$ field which   receives its mass through the interaction  with topological charges $Q$.   
  Uncovering this   picture (which allows us to measure the topological charges of constituents) was the main motivation 
  for introducing the chiral condensate into the theory.\\
 {\bf h.} We should also remark here that a similar picture  for the instanton interactions 
occurs at large chemical potential   $\mu\gg \Lambda_{QCD}$ in deconfined, the so-called color superconducting phase  \cite{ssz1}. In the present case $T>T_c$ the weak coupling regime (small instanton density ) is governed  by small parameter $  a\sim \exp(-\gamma N)\ll 1$ while 
at large $\mu\gg \Lambda_{QCD}, N=3$ case  
the corresponding small factor is $\left(\Lambda_{QCD}/\mu\right)^b\ll 1$\cite{ssz1}. \\

\section{ Small $T<T_c$: Confined phase. Speculations. } 
In this section 
we want to speculate on  the fate of  the instantons when we cross the phase transition line at $T=T_c$. 
To be more precise: we want  to see if any traces of well defined   instantons    discussed above
 can be recovered.  The instanton expansion is not justified
 in  the strong coupling regime $T<T_c$ where the expansion coefficient becomes of order one, $a\sim 1$.

 Therefore, we  do not even attempt to use instanton calculus  or any other semiclassical computations in the present section.  Instead, we present  a few  indirect arguments supporting the picture that the instantons   do not completely  disappear from the system when we cross the phase transition line from above, but rather dissociate  into the instanton quarks \cite{Fateev,Belavin}, 
  the self-dual objects with fractional topological charges $\pm 1/N$
which become the dominant quasi-particles. The arguments are not based on the semiclassical 
calculations, but rather on analysis of the low energy lagrangian written  in the dual form similar 
to CG representation discussed in the previous section. Our proposal about the  fate of  instantons at $T<T_c$ originally derived  in ref.\cite{JZ} and to be reviewed for completeness in this section  
should be considered as one of the many possible outcomes. In this sense this section is very speculative in  nature in contrast with previous sections where
weak coupling regime is justified at large $N$ and precise statements can be made. 
 
 We start from the chiral Lagrangian and keep only the singlet  $\eta'$ field.
  We assume the following expression for the effective Lagrangian for  the $\varphi$  field
  which has a specific Sine-Gordon (SG) form, 
 \bea
 \label{chiral}
 L_{\varphi}= \frac{1}{2}f_{\eta'}^2 
( \partial_{\mu} \varphi)^2 + E_{vac} \cos\left( \frac{ \varphi - \theta }{N}\right), ~~~~
f_{\eta'}^2 m_{\eta'}^2=\frac{E_{vac}}{N^2}
 \eea
 where $E_{vac} \sim N^2$ is the vacuum energy of the ground state in the chiral limit,
 expressed in terms of the gluon condensate,
 \beq
 \label{E}
 E_{vac}= \frac{1}{4} \la \Theta^{\mu}_{\mu}\ra=\la b
\alpha_s /(32 \pi) G^2 \ra \sim N^2,
 \eeq
 where we use the standard  expression for the conformal anomaly of the  energy -momentum tensor, $\Theta^{\mu}_{\mu}$. The expression (\ref{chiral}) of course satisfies the standard requirement
 crucial for the resolution of $U_A(1)$ problem: the vacuum energy in gluodynamics depends on $\theta$ through the combination $\theta/N$. It also has a
    very specific SG  structure for the singlet combination  
  corresponding to the following behavior of the $(2k)^{\rm th}$
   derivative of the vacuum energy in pure gluodynamics \cite{veneziano},
\bea
\label{veneziano}
\frac{ \partial^{2k} E_{vac}(\theta)}{ \partial \, \theta^{2k}}|_{\theta=0} \sim \int \prod_{i=1}^{2k} dx_i \langle
Q(x_1)...Q(x_{2k})\rangle  
\sim (\frac{i}{N})^{2k}, ~~~~~~~ {\rm where}~~
Q\equiv\frac{g^2}{32\pi^2} G_{\mu\nu} {\widetilde G}_{\mu\nu}.
\eea
The same structure was also advocated in \cite{HZ} from a different
perspective. We shall not discuss any  additional arguments supporting  such Sine-Gordon structure
referring to the original papers\footnote{One more additional argument supporting SG structure  
 $\sim \cos\left( \frac{\theta }{N}\right) $ in pure gluodynamics will be given later in the text.}. This is precisely the place where the term `` speculation" from the title of this section, enters our analysis. One should also note that the combination 
 \beq 
 \chi_g = \frac{E_{vac}}{N^2} =\frac{ \partial^{2} E_{vac}(\theta)}{ \partial \, \theta^{2}}|_{\theta=0}
 \nn
 \eeq 
 is nothing but topological susceptibility $\chi_g$ for gluodynamics in the large $N$ limit.

 Now we want to represent the low energy lagrangian (\ref{chiral}) in the dual form (CG representation) to see
 if any traces from the instantons discussed at $T>T_c$ can be recognized. The effective lagrangian 
 is obviously the color singlet object.  Therefore, all color dynamics can not be recovered by this method.
 However, the topological charge is color singlet operator which is coupled to $\theta$.
 The $\theta$ parameter is not a dynamical field in QCD, however the $\eta'$ field is, and it   always enters   the dynamics in combination $(\theta -\varphi)$. Let us repeat again that this was the main reason to introduce the   chiral condensate into the sysytem:
 it allows to study the dynamics of the topological charges. Therefore, in principle, the analysis 
 of the $\eta'$ field gives the information about the topological charges of the constituents.
 We use the trick (SG-CG mapping) below  to attempt to answer the following question: what kind of constituents 
can provide the low energy behavior (\ref{chiral},\ref{veneziano})?
 
 We use the technique developed in the previous section and represent SG action in the dual form.
 Technically, it goes as follows:  eq. (\ref{chiral}) replaces expression (\ref{cos}) discussed previously.   
 \exclude{
 The standard Witten-Veneziano formula is obtained by differentiating vacuum energy with respect to theta,
 \beq
\label{5}
\lim_{ q \rightarrow 0} \; 
i \int dx \, e^{iqx} \la T \left\{ \frac{\alpha_s}{8 \pi} 
G \tilde{G} (x)  \, 
\frac{\alpha_s}{8 \pi} G \tilde{G} (0) \right\}  \ra =  
- \frac{ \partial^{2} E_{vac}(\theta)}{ \partial \, \theta^{2}} 
\eeq
}
 As in (\ref{cos}) the Sine-Gordon effective field
 theory (\ref{chiral}) can be represented in terms 
 of a classical statistical ensemble (CG representation) similar to
 (\ref{CG})  with the replacements $\lambda\rightarrow E_{vac}$, $d^3x\rightarrow d^4x$ as we assume 
 zero temperature $T=0$ in this phase.
 By repeating all previous steps we arrive at the following expression
 \bea
\label{CG1}
Z =  
\sum_{M =0}^\infty \frac{(\frac{E_{vac}}{2})^M}{M! }  \int d^4x_1 \dots \int d^4x_M \cdot
\sum_{Q_a =\pm 1/N}\int D\varphi e^{-\frac{1}{2}f_{\eta'}^2\int d^4x( \partial_{\mu} \varphi)^2}\cdot
 \left(e^{i \sum_{a=1}^M Q_a\left[\varphi(x_a)-\theta\right] }\right) . 
 \eea
The functional integral is trivial to perform and one arrives at the dual CG action,
\bea
\label{CG2}
Z=\sum_{M_\pm=0}^\infty \frac{(\frac{E_{vac}}{2})^M}{M_+!M_-!}  
\int d^4x_1 \ldots \int d^4x_M\cdot
e^{-i\theta\sum_{(a=0,Q_a =\pm 1/N)}^M Q_a }\cdot e^{-{1\over f_{\eta'}^2}\sum_{(a>b=0, Q_a =\pm 1/N)}  Q_aG(x_a-x_b)Q_b
} ,
\eea
where $ G(x_a-x_b)$ is the $4d$ Green's function,
\beq
\label{G-4}
G(x_a - x_b) = {1\over 4\pi^2 ({x}_a-{x}_b)^2}.
\eeq
 The fundamental difference in comparison with the previous case
 (\ref{CG}) is that
while the total charge is integer, the individual charges are
   { \bf fractional $\pm 1/N$}.
This  is a
direct   consequence of the $\theta/N$ dependence in the underlying
effective Lagrangian (\ref{chiral}) before integrating out $\varphi$ fields, see eq. (\ref{CG1}).

 A few remarks on 
 physical interpretation of the  CG representation (\ref{CG2}) of
 theory (\ref{chiral}) are in order:\\  
 {\bf a.} As before,  one can identify
$Q_{\rm net}\equiv \sum_a Q_a$ with  the   total topological charge of
the given configuration.\\ 
{\bf b.} Due to the $2\pi$
periodicity of the theory, only configurations which contain an integer
topological number contribute to the partition function. Therefore, 
the number of particles for each given configuration $Q_i$ with
charges $ \sim 1/N$  
 must be proportional to $N$.\\
{\bf c. } Therefore, the number of integrations
over $d^4x_i$   in CG representation exactly equals $4 N k$, where 
$k$ is integer.   This number  $4 N k$ exactly
corresponds to the number of zero modes in the  $k$-instanton
background. This is basis for the conjecture \cite{JZ}  that 
at low energies (large distances) the fractionally charged species,
$Q_i=\pm 1/N$   
are the {\bf   instanton-quarks} suspected long ago \cite{Fateev}.\\
{\bf d.} For  the gauge group, $G$
the number of integrations would be equal to  $4k C_2(G)$  where
$C_2(G)$ is the quadratic Casimir of the gauge group
  ($\theta$ dependence in physical observables comes  in
the combination $\frac{\theta}{C_2(G)}$).   This number  $4k C_2(G)$
exactly corresponds 
to the  number of zero modes  in the $k$-instanton background for gauge
group $G$.\\
{\bf e.}  We do not use the weak coupling regime or instanton calculus anywhere  in our arguments. 
Still, we recover the moduli space which we identify with strongly interacting instantons
in the confinement  phase of the theory.\\
{\bf f.} Role of the fugacity for this statistical ensemble plays $E_{vac}\sim N^2$.
Therefore, an average distance between constituents is of order $\bar{r}\sim E_{vac}^{-1/4}\sim 
\Lambda_{QCD}^{-1}N^{-1/2}$ which suggests that the system is very dense.
It obviously implies that the instanton expansion makes no sense in this regime
as all terms are equally important, which is in huge contrast with hierarchy
from the  previous case at $T>T_c$, (\ref{hierarchy}). \\
{\bf g.}
The Debye screening length $r_{D}\sim m_{\eta'}^{-1}\sim \Lambda_{QCD}^{-1}\sqrt{N}\gg \bar{r}$ is   large.  It means that the number of constituents participating in the screening is  order of
$(r_D/\bar{r})^4 \sim N^4$. \\
{\bf h.} According to eq. (\ref{chiral}) the number of  instanton quarks 
in the spacetime box of size $\Lambda_{QCD}$ should be $N^2$
as an average distance between constituents is $\bar{r}\sim N^{-1/2}$. 
Each instanton contains $N$ instanton quarks, hence
the density  of instantons  should be of order $N \Lambda_{QCD}^4$.\footnote{
 In \cite{JZ} it was conjectured that these constituents (instanton quarks) are the driving force for the confinement.}
It is consistent with observation from holography, section II that any finite number of instantons
will disappear from the system.
\exclude{{\bf i.} One obvious  objection for an identification  of $Q_a$ with the
topological charge
immediately comes in mind: 
  it has long been known  that   instantons 
can explain most low energy QCD phenomenology \cite{ILM}   with the
 exception   confinement;  and we claim
that   confinement also arises in this picture: how can this be consistent?   
 We note that quark confinement can not be described in the  dilute gas 
approximation, when the instantons and anti-instantons are well
separated and maintain their individual properties (sizes, positions,
orientations), as it happens at large $T>T_c$, see previous section.  However,   
in strongly coupled theories the instantons and
anti-instantons lose their individual properties (instantons will
``dissociate '')  their sizes become very large and they overlap.  The relevant
description is that of instanton-quarks which carry the magnetic charges
and  which can propagate  far away from
instantons- parents  being  strongly correlated with each other. For such  
configurations the confinement
is a possible outcome of the dynamics.\\}

\subsection{ The relation to other studies} 
 As we mentioned above our arguments in the present section look extremely speculative as they are not based
 on instanton calculus or any other dynamical calculations which include
   color degrees of freedom.  Still, by analyzing $\theta$ dependence in deconfining phase
    we infer (indirectly)   that some fractionally charged degrees of freedom emerge
    at low $T<T_c$. We presented arguments suggesting that the corresponding fractionally charged
    constituents are the colored instanton quarks.   
      We should remark here that the fractionally charged constituents have been discussed 
     in the literature  in a number of papers previously. In particular, there seems
 to be a close relation between     
instanton quarks and the ``periodic instantons" 
\cite{vanbaal,diakonov,Gattringer}, center vortices and nexuses with fractional fluxes $1/N$, see  e.g.\cite{Engelhardt} and references therein. We shall not  discuss the corresponding 
connections in details in the present paper by referring to the original literature and the recent review by one of the author \cite{Zhitnitsky:2006sr} where  some comments on the corresponding connections
have been made.
In the present work  we want to make a few comments on two recent papers  \cite{Unsal:2008ch}, \cite{Diakonov:2007nv} where 
the picture similar to the one presented  in this work is advocated. 

We start with \cite{Unsal:2008ch}.  In that  work the authors consider a 
specifically deformed $SU(N)$ gluodynamics at $T\neq 0$.
It has been shown that such a deformation  
supports a  reliable analysis  in the weak coupling regime in the confining phase. The results of the corresponding calculations imply
that the relevant degrees of freedom in the confined phase are the 
self dual magnetic monopoles with action $\frac{8\pi^2}{g^2N}$
and  topological charges $Q=\pm 1/N$ which are precisely the features of the instanton quarks 
discussed above. 

In contrast, the starting point of ref.\cite{Diakonov:2007nv} is  semiclassical calculations in the background  of calorons\cite{vanbaal} where the weak coupling regime can not be guaranteed.  While the calculations are semiclassical in nature, and therefore, can not be trusted in the strong coupling regime, still, the corresponding analysis   
shows how well localized instantons with integer topological charges at $T>T_c$  
may dissociate into the fractional constituents at $T<T_c$, and become the key players
in the confining phase. This is precisely the picture we are advocating in the present work
based on analysis of sharp $\theta$ changes at $N=\infty$.  It is impressive how
complicated  semiclassical calculations carried out in \cite{Diakonov:2007nv} 
lead to the expression for the vacuum energy $E_{vac}\cos(\frac{\theta}{N})$ advocated in \cite{HZ}
using completely different technique. 
\footnote{Such a SG structure was a  crucial element 
  for recovering the   fractional topological charges $Q=\pm 1/N$ in the confining phase 
  using $\eta'$ as a probe, see section V and original discussions in \cite{JZ}.}
 Our technique does not allow us to make any dynamical calculations in this phase 
  as all color degrees of freedom  have been integrated out in the course of  obtaining (\ref{chiral}). In other 
words, we can not study the dynamics   
 of fractionally charged constituents in contrast with  papers \cite{Unsal:2008ch, Diakonov:2007nv}
 \footnote{In particular, we do not see a beautiful picture of a multi-component   color  Coulomb 
plasma with nearest-neighbor   interactions in the Dynkin space advocated in \cite{Unsal:2008ch, Diakonov:2007nv}. 
 Still, we do see the color- singlet Coulomb interaction of the fractionally   charged   $\pm 1/N$ constituents due to 
$\eta'$ at very large distances where color already confined. }. 
 However, the fact that the constituents carry fractional topological charge $1/N$
  can be recovered  in our approach because the color- singlet 
 $\eta'$ field enters the effective lagrangian as    $ \cos(\frac{\theta-\varphi}{N})$ and serves as a perfect probe   of the  topological charges of the constituents. 
One should also emphasize that the procedure of the recovering of the fractional  topological charge $1/N$ (which has been used here) is not based on the weak coupling expansion. 
 
Our final comment in this subsection is as follows.
The main ingredient  in holographic picture discussed in section II
was D0 brane wrapping around $x^4$ which  behaves differently in confined and 
deconfined phases, and correspondingly leads to a different $\theta$ behavior
on opposite sides of phase transition line.
Similar picture was also observed  in ref.\cite{Gorsky:2007bi} where the authors studied
the D2 brane in confined and deconfined phases
to arrive to the same conclusion on sharp changes in $\theta$ behavior.
 The topological objects (sensitive to $\theta$)
were identified as magnetic strings in ref.\cite{Gorsky:2007bi}. These objects  apparently have been observed in the lattice simulations\cite{Chernodub:2007rn}.

\section{conclusion } 

We 
explore the consequences of the assumption that in the large $N$ QCD 
and QCD-like theories 
confinement-deconfinement phase transition
takes place at the  temperature where the dilute instanton
calculation breaks down, and $\theta$ dependence drastically changes.  
This assumption 
is supported by holographic and field theoretic arguments.
At very high temperatures,  $T\gg \Lambda_{QCD}$ instantons are   well localized configurations 
with a typical size $\rho\sim T^{-1}\ll \Lambda_{QCD}^{-1}$. 
As the temperature is getting lower  the  instanton size becomes 
of order $\rho\sim \Lambda_{QCD}^{-1} $ however provided the perturbative corrections
in the instanton background do not significantly change the picture, the instanton density remains dilute.
Instanton expansion breaks down at $T_c$, and 
for $T<T_c$, the strong interacting regime and confinement are realized. 
Instantons are no longer well localized configurations for $T<T_c$, but rather,
in the picture of Section V, they  are represented by
 $N$ instanton quarks which can propagate far away from each other.  
The presence of the light field $\eta'$ in the model
is important for this picture and a specific lagrangian is assumed in Section V.
The mass of $\eta'$ in both phases 
 in the dual picture can be thought as the Debye screening mass generated by the 
Coulomb interaction of the topological charges.  

We have made a number of assumptions in the field theoretic analysis to 
arrive at the conclusion that the $\theta$ dependence changes sharply at some
value of $T_c$.
We also used the holographic model to argue that this transition coincides
with confinement-deconfinement phase transition.
This conclusion is supported by the lattice simulations  \cite{Alles:1996nm,Lucini:2003zr,Lucini:2004yh,Del Debbio:2004rw,Lucini:2005vg}.
  
The value of
the critical temperature  as a function of (sufficiently small) chemical potential 
$T_c(\mu)$ is estimated in Section III. The obtained expression  is in excellent agreement  
with numerical computations  \cite{de Forcrand:2002ci,de Forcrand:2003hx,mu}.
Finally,  we presented the arguments  that this line of the phase transition 
(which is the first order for large $N$ and $N_f\ll N$)
continuously  transforms into the line $\mu_c(T)$ studied at large $\mu$
in ref.\cite{Zhitnitsky:2008ha}.
The argument is based on the observation that  the physical nature  of the phase transition
along entire line is the same:
it is the drastic changes in $\theta$ behavior when the phase transition line is crossed.

  
It would be very interesting to see if Coulomb law 
between instantons can be understood holographically in the deconfining phase. 
As we mentioned in the text
it is quite nontrivial \exclude{result} that at large distances in the presence of $\eta'$ field the interaction between instantons and anti instantons
is the same (up to sign) as the interaction between two instantons. 

Finally, we would like to make a short comment on relevance of the present analysis
to real QCD with $N_f=N=3$.  The main subject of the present paper is pure gluodynamics at large $N$,
and therefore one can not immediately  apply the results of the present analysis to the
real QCD with $N_f=N=3$. However, we do expect that our results can be and should be compared with the lattice simulations for pure gluodynamics for $N\geq 3$ and for QCD with $N_f\ll N$ when the first order phase transition is expected, see \cite{Alles:1996nm,Lucini:2003zr,Lucini:2004yh,Del Debbio:2004rw,Lucini:2005vg}.  For the case $N_f\sim N$ our technique is not applicable as we explained
in section IIIB. In this case one should expect that the properties of the phase transition is very sensitive
to the details of the fermion matter fields. It a subject of a separate analysis which  will not be discussed here. However, 
we expect that the picture of the phase transition as a transition between plasma phase
(when the instanton quarks are in plasma state at $T< T_c$) and molecular phase (when the instanton quarks form a small instanton at $T>T_c$) qualitatively describes real QCD with $N=3$ as the confinement in non- abelian gauge theories is  determined by the dynamics  of gluon (not quark)  degrees of freedom.

This work 
was supported, in part, by the Natural Sciences and Engineering
Research Council of Canada.   
We thank   the KITP, Santa Barbara
for organizing the workshop
 "Nonequilibrium Dynamics in Particle Physics and Cosmology" 
 where this work was initiated.
We also  thank The Galileo Galilei Institute for Theoretical Physics and
the organizers of the workshop ``Non-perturbative methods in strongly coupled
gauge theories'' for hospitality while this work was completed.
 We thank E. Shuryak, M. Teper, M. Unsal and L. Yaffe for discussions and comments on the manuscript.
We also thank  M. Unsal and L. Yaffe  for
  providing \cite{Unsal:2008ch}
 prior to publication.  This research was supported in part by the National Science Foundation under Grant No. PHY05-51164.

\end{document}